\documentclass{article}

\usepackage{arxiv}

\usepackage[utf8]{inputenc} % allow utf-8 input
\usepackage[T1]{fontenc}    % use 8-bit T1 fonts
\usepackage{url}            % simple URL typesetting
\usepackage[colorlinks=true,citecolor=blue]{hyperref}
\usepackage{booktabs}       % professional-quality tables
\usepackage{amsfonts}       % blackboard math symbols
\usepackage{nicefrac}       % compact symbols for 1/2, etc.
\usepackage{microtype}      % microtypography
\usepackage{lipsum}
\usepackage{physics}
\usepackage{amsmath}
\usepackage{amssymb}
\usepackage{amsthm}
\usepackage{stackrel}
\usepackage{comment}
\usepackage{todonotes}
\usepackage{graphicx}
\usepackage{color}
\usepackage{enumitem}
\usepackage{algorithm}
\usepackage[noend]{algpseudocode}

\usepackage{caption}
\usepackage{float}
\usepackage{placeins}
\usepackage{multirow}

\newlist{todolist}{itemize}{2}
\setlist[todolist]{label=$\square$}
\usepackage{pifont}

\definecolor{orcidlogocol}{HTML}{A6CE39}

\title{Hypothesis Testing for Error Mitigation: \\How to Evaluate Error Mitigation}

\author{	\href{https://orcid.org/0000-0002-6597-2770}{\includegraphics[scale=0.06]{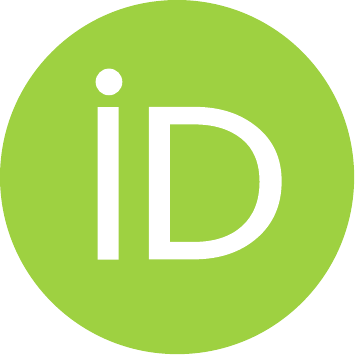}\hspace{1mm}Abdullah Ash Saki}  \\
	Zapata Computing, Inc.\\
	100 Federal Street, 20th Floor\\
	Boston, MA 02110 \\
	\texttt{saki@zapatacomputing.com} \\
	\And
	\href{https://orcid.org/0000-0002-0078-5202}{\includegraphics[scale=0.06]{orcid.pdf}\hspace{1mm}Amara Katabarwa} \\
	Zapata Computing, Inc.\\
	100 Federal Street, 20th Floor\\
	Boston, MA 02110,\\
	\texttt{amara@zapatcomputing.com} \\
	\AND
	\href{https://orcid.org/0000-0002-9050-3685}{\includegraphics[scale=0.06]{orcid.pdf}\hspace{1mm}Salonik Resch} \\
	Zapata Computing, Inc.\\
	100 Federal Street, 20th Floor\\
	Boston, MA 02110,\\
	\texttt{Salonik.Resch@zapatcomputing.com} \\
        \And 
	\href{https://orcid.org/0000-0003-4376-2147}{\includegraphics[scale=0.06]{orcid.pdf}\hspace{1mm}George Umbrarescu} \\
	Department of Physics and Astronomy\\
        University College London\\
	London, United Kingdom\\
	\texttt{george.umbrarescu.20@ucl.ac.uk} \\
}

\begin{document}
\maketitle
\begin{abstract}
In the noisy intermediate-scale quantum (NISQ) era, quantum error mitigation will be a necessary tool to extract useful performance out of quantum devices. However, there is a big gap between the noise models often assumed by error mitigation techniques and the actual noise on quantum devices. As a consequence, there arises a gap between the theoretical expectations of the techniques and their everyday performance.  Cloud users of quantum devices in particular, who often take the devices as they are, feel this gap the most. How should they parametrize their uncertainty in the usefulness of these techniques and be able to make judgement calls between resources required to implement error mitigation and the accuracy required at the algorithmic level? To answer the first question, we introduce hypothesis testing within the framework of quantum error mitigation and for the second question, we propose an inclusive figure of merit  that accounts for both resource requirement and mitigation efficiency of an error mitigation implementation. The figure of merit is useful to weigh the trade-offs between the scalability and accuracy of various error mitigation methods. Finally, using the hypothesis testing and the figure of merit, we experimentally evaluate $16$ error mitigation pipelines composed of singular methods such as zero noise extrapolation, randomized compilation, measurement error mitigation, dynamical decoupling, and mitigation with estimation circuits. In total our data  involved running $275,640$ circuits on two IBM quantum computers.
\end{abstract}

% keywords can be removed

\section{Introduction}
The current state of quantum computers is often dubbed as the noisy intermediate-scale quantum (NISQ)~\cite{Preskill2018} regime. 
In this age, qubit count and qubit and gate quality still need to be improved before quantum error correction can be done successfully. Nonetheless, it is a range in which it should be possible to do computations that cannot efficiently be simulated on a classical computer. Since its inception there has been a burst of research looking for a quantum advantage in different areas like quantum machine learning \cite{Alcazar2020, Schuld2020, Schuld2021, Cao2021, Liu2018, Benedetti2019, Perdomo-Ortiz2018, Rudolph2020b}, quantum chemistry~\cite{qchem1,qchem2,qchem3,qchem4,qchem5, cao2019quantum}, and quantum finance \cite{Rebentrost2018, Woerner2019, Alcazar2022}. An excellent and comprehensive view of near-term quantum algorithms is contained here \cite{Bharti2021}. As one would expect, alongside this flurry of research on the algorithmic side arose the field of quantum error mitigation. 
Quantum error mitigation first rose in the ideas of Richardson Extrapolation and Probabilistic Error Cancellation (PEC) \cite{Temme2017a}. In the first method, meant to correct the expectation value of the operator $E$, the user runs versions of the unmitigated circuit with increasing noise levels, which is done by stretching the gate times. This gives one a set  $ \mathcal{E} = \{  \langle E \rangle_{1}, \langle E \rangle_{2} \dots  \langle E \rangle_{i} \dots \langle E \rangle_{M} \}  $,  where for each increasing $i$,  $\langle E \rangle_{i}$ was estimated using a circuit with longer gate times. One then extrapolates to the zero noise limit using the Richardson extrapolation technique borrowed from solving differential equations. In the second method, one does a careful tomographic characterization of some basis gates $\mathcal{O}_{j\alpha}$ for one's device, where $j$ is a label for a gate $\mathcal{G}_j$ in the unmitigated circuit $\mathcal{C}$, while $\alpha$ will be a label indexing the linear combination for gate $\mathcal{G}_j$. Thinking about $\mathcal{C}$ this way means that $\mathcal{C}$ has been replaced with an ensemble of circuits $\{ \mathcal{C}^{(k)} \}$ for which we will sample from with the selected circuit run on the quantum device. It was shown that this gives an unbiased estimate of the observable we would like to estimate.   These techniques were experimentally demonstrated soon after their invention~\cite{Kandala2019a}. While these techniques were foundational to error mitigation, they had/have considerable barriers to the typical user of near-term quantum devices who receives access to quantum devices through the cloud. The first method requires pulse-level access which is not easily obtainable, while the second requires process tomography, which is a considerable overhead for a cloud user with limited access time. 
Motivated by the limitations of pulse-level access, a slew of works were produced investigating a digitized version of Richardson Extrapolation \cite{Giurgica-Tiron2020a, Pascuzzi2022, Urbanek2021}, while for PEC, recent work has been done to reduce the overhead of quantum tomography by combing it with cycle benchmarking and randomized compiling \cite{Berg2022}. The digitized versions of ZNE have an ambiguity as to how to do the extrapolations, i.e., what functions to choose for real devices; solutions borrowing ideas from machine learning have been developed to work around this problem \cite{Czarnik2021}, namely Clifford Data Regression (CDR). Another exciting idea for error mitigation comes roughly from thinking of a spatialized version of the quantum Zeno effect where one considers copies of the noisy state \cite{Huggins2021a}, called Virtual State Distillation (VSD). There have been a couple of attempts to combine different error mitigation techniques to overcome the specific limitations of a single error mitigation algorithm; so \cite{Lowe2020, Bultrini2021} came up with a framework that combines  ZNE, CDR, and VSD, while \cite{Mari2021} proposed to combine PEC and ZNE. These are all general-purpose techniques, but one can imagine using information about the problem in hand, for example, symmetries one expects a unitary to have to mitigate errors; \cite{McArdle2019a, Bonet-Monroig2018, Cai2021} have developed ideas along these lines.

As we proceed deeper into the NISQ era, a few questions need to be understood and answered.

\begin{enumerate}[label=(\roman*)]
    \item What is the precise asymptotic scaling of resources requirements \cite{Tsubouchi2022, Takagi2022}? The resource scaling will be critical as we incorporate some quantum error correction with quantum error mitigation \cite{Piveteau, Lostaglio2021, Bultrini2022}.
    
    \item How can one easily implement and benchmark these methods for everyday use? The \texttt{Mitiq} \cite{LaRose2020} and \texttt{Qermit}\cite{Cirstoiu2022} software packages are efforts in this direction. 
    
    \item Often, the methods make assumptions about the quantum noise like Markovianity, locality of noise in the operations, and time independence. These assumptions mean that it is unclear whether these methods work in everyday use and how an actual experiment on the cloud will perform. Therefore, it seems plausible that a sequence of quantum error mitigation techniques might need to be designed for a specific hardware and quantum algorithm. How then can we quantify our uncertainty and ensure that particular sequence is not only improving our results, but has been chosen in such a way that limited resources have been used in a near-optimal fashion?
\end{enumerate}

Our work is dedicated to answering the last question. For this, we introduce into quantum error mitigation a notion of hypothesis testing for quantifying our confidence in different \textit{error mitigation pipelines}, where an \textit{error mitigation pipeline} is a single or a sequence of multiple error mitigation (EM) techniques.

We make the following contributions in this paper:
\begin{enumerate}[label=(\roman*)]
    \item With a series of mitigation techniques in hand, all with their limitations and a desire to combine them, how can one evaluate which set or subset of techniques is responsible for mitigation? This is crucial since resources may be limited. For this question, we initiate the use of hypothesis testing within the framework of error mitigation.
    
    \item Different error mitigation strategies produce overhead in different ways, i.e., increasing the number of distinct circuits one needs to run, increasing the number of shots one needs to run for any particular circuit and lastly increasing the depths of the circuit. For this issue we propose an entropic figure of merit that considers the different ways the overhead can appear.
\end{enumerate}

The rest of the paper is organized as follows: Sec.~\ref{sec:hypothesis-testing} introduces the hypothesis testing framework used in this paper to evaluate error mitigation (EM) pipelines. In Sec.~\ref{sec:em-details}, we discuss the details of error mitigation pipeline constructions and hardware experiments. Accuracy and resource trade-off considerations are described in Sec.~\ref{sec:metric} along with metrics encapsulating them. We present the experimental results in Sec.~\ref{sec:results} and finally, we conclude in Sec.~\ref{sec:conclusion}.

\section{Hypothesis Testing: Modeling our uncertainty about error mitigation}\label{sec:hypothesis-testing}
As a more and more cloud quantum computers come online, it will be necessary to do quantum error mitigation. On the other hand, the various complicated noise models on different platforms reduce the efficacy of error mitigation. This raises  important questions: Is error mitigation actually working ? How typical or representative are my conclusions about the efficacy of error mitigation for a specific device and how certain can I be? 
There are many EM techniques on hand; we can choose a single method or combine any number to mitigate errors in hardware experiments. For our work, we choose measurement error mitigation (\texttt{MEM}), randomized compiling (\texttt{RC}), zero noise extrapolation (\texttt{ZNE}), and dynamical decoupling (\texttt{DD}). However, we believe our ideas will be vital as the complexity of techniques increases when using other techniques. Since there are different versions of \texttt{ZNE} depending on the extrapolation method and the folding method, we introduce the following notation: $\texttt{ZNE}^{(E)}$,  where the presence of $E$ represents whether \textit{estimation circuits}~\cite{Urbanek2021} were used. Following Cirstoiu et al. in~\cite{Cirstoiu2022}, we use the relative error mitigation ($REM$) metric to characterize the performance of an experiment, defined as 

\begin{equation}\label{eq:rem}
    \text{REM} = \frac{|\langle E \rangle_{ideal} - \langle E \rangle_{mitigated}|}{|\langle E \rangle_{ideal} - \langle E \rangle_{noisy}|}
\end{equation}

It measures how close a mitigated expectation value $(\langle E \rangle_{mitigated})$ is to the ideal expectation value $(\langle E \rangle_{ideal})$ compared to a noisy (unmitigated) expectation value $(\langle E \rangle_{noisy})$. A $REM$ $<1$ means the EM pipeline is mitigating the errors and pushing the mitigated expectation value closer to the ideal value compared to the unmitigated noisy value. On the other hand, a $REM$ $\geq 1$ indicates that the error mitigation is making the expectation value worse. 

Another crucial theme will be accuracy vs resource trade offs. To motivate why this might be an interesting problem to contemplate, consider the following results clipped from Table~\ref{tab:perth-local}:

\begin{table}[H]
    \centering
    \begin{tabular}{ccc}
      \hline
      Pipeline & $REM$ & $REM$  \\
         & ($p=1$)& ($p=2$) \\ \hline
       $\mathcal{P}_8 $ &  0.49   & 0.30   \\
        $\mathcal{P}_4$ & 0.60 & 0.30  \\ \hline
    \end{tabular}
    \caption{Data from \texttt{ibm\_perth} showing relative error mitigation ($REM$)}
    \label{tab:pipeline_motivation}
\end{table}
here $\mathcal{P}_4$ is a pipeline consisting of $\texttt{ZNE} + \texttt{DD} + \texttt{MEM},  $ while $\mathcal{P}_8$ is a pipeline consisting of $\texttt{ZNE} + \texttt{RC} + \texttt{DD} + \texttt{MEM} $. This is a rather simple case of what could happen generally, i.e. investing more resources may not yield better results.  We shall see that, in this case, $\mathcal{P}_8$ requires resources roughly $2.5$ times that of $\mathcal{P}_4$, yet by choosing the right extrapolation function in $\texttt{ZNE}$ on the right device, resources can be saved. This is a rather obvious trade-off to make but there will be situations where more nuanced judgement calls will need to be made. We are therefore offering tools for this analysis.

\begin{figure}[t]
    \centering
    \includegraphics[height=120mm]{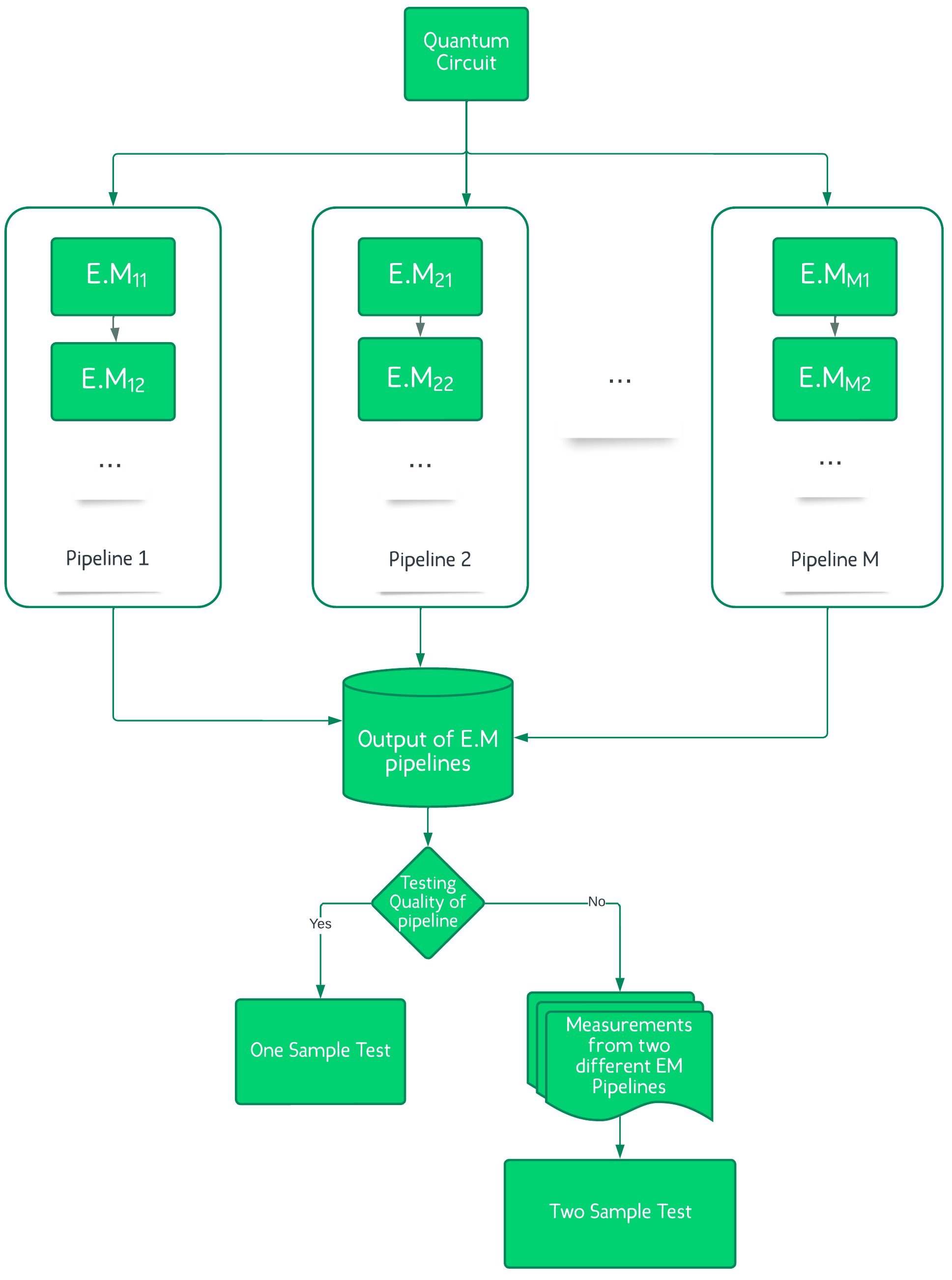}
    \caption{Hypothesis testing for quantum error mitigation.}
    \label{fig:flowchart}
\end{figure}

We propose using statistical hypothesis testing. Specifically, we use \emph{one-sample test of proportions} to answer how good a pipeline is and \emph{two-sample test of proportions} to compare two different pipelines. We summarize the procedure in Figure~\ref{fig:flowchart}.

We want to emphasize that this sort of analysis or variant thereof is a necessary precursor to any ideas related to the application of volumetric benchmarking \cite{BlumeK2019, Lubinski2021, Hamilton2022, Miller2022}
, for the simple reason that in practice a user will need to do quantum error mitigation for experiments and is interested in the performance of a quantum device in this context. It behooves the user to make sure that the techniques being used have been properly chosen for the device at hand.

\subsection{One-sample test of proportions}\label{sec:one-sample-test}
\textit{One-sample test of proportions} can determine if one outcome is more likely to happen than other in a binomial distribution. We employ this test to understand if an error mitigation pipeline succeeds (mitigating errors) more often than it fails. To decide whether an error mitigation attempt is successful or not, we use the concept of $REM$~\cite{Cirstoiu2022} (Eq.~\ref{eq:rem}). We label the experiments with $REM$ $< 1$ as \texttt{SUCCESS} and experiments with $REM$ $\geq 1$ as \texttt{FAIL}, and thus convert them to binary values for the one-sample test of proportions.

Next, we construct the following null and alternate hypotheses to determine if \texttt{SUCCESS} is occurring significantly more than \texttt{FAIL}:

\begin{itemize}
    \item Null hypothesis, $H_0: \hat{p} = p_0$ with $p_0 = 0.5$.
    
    \item Alternative hypothesis, $H_A: \hat{p} > p_0$.
\end{itemize}

Here, $\hat{p}$ is the proportion of successful experiments, which is computed as
\begin{equation}
    \hat{p} = \frac{ \text{ \# of successful experiments}}{ \text{ \# total experiments }(n)},
\end{equation}
and $p_0$ is the known proportion. A value of $0.5$ specifies that the EM pipeline is equally likely to succeed and fail, i.e., the error mitigation technique is indistinguishable from the case of no error mitigation. 

The test statistics are computed using the following formula:
\begin{equation}
    z^*_{(one-sample)} = \frac{\hat{p} - p_0}{\sqrt{\frac{p_0 (1 - p_0)}{n}}}
\end{equation}

Based on the test statistics, the null hypothesis can or cannot be rejected. If the null hypothesis cannot be rejected, it will mean the EM pipeline is likely to generate random results, i.e., ineffective. On the other hand, if the null hypothesis is rejected it tells us that the pipeline mitigates errors more often than it fails.

\subsubsection*{Confidence interval: one-sample proportions}\label{sec:conf-int-one-sample}
Having decided whether or not to accept or reject the null hypothesis, this framework allows us to attach a confidence level to our conclusion. For this work we shall compute the $95\%$ confidence interval of the proportion of the successful experiments, $\hat{p}$ as follows:
\begin{equation}\label{eq:ci-one-sample}
    CI = 100 \times (\hat{p} \pm z^* \times SE)
\end{equation}

where, standard error, $SE = \sqrt{\hat{p}(1-\hat{p})/n}$ and $z^* = 1.96$ for $95\%$ confidence level.

Suppose we get $CI = 80\% \pm 10\%$ for a pipeline $\mathcal{P_A}$. It tells us that the pipeline successfully mitigates errors (i.e., generates $REM$ $< 1$) $70\%$ to $90\%$ of experiments. A user might be interested in pipelines with a higher confidence interval of successful trials.

\subsection{Two-sample test of proportions}\label{sec:two-sample-test}
As discussed before, we are particularly interested in comparing two different EM pipelines. For that purpose, the \textit{two-sample tests of proportions} which can determine whether the two populations differ significantly on a specific characteristic will be used. In our proposed framework, two populations are two sets of experiments, each with a different error mitigation (EM) pipeline, such as one with only \texttt{ZNE} vs.  \texttt{ZNE} + \texttt{RC} +  \texttt{DD} + \texttt{MEM}. As the \emph{specific characteristic}, we again pick the $REM$~\cite{Cirstoiu2022} value.
 
Let $\hat{p}_A$ and $\hat{p}_B$ represent proportions of success for two EM pipelines, $\mathcal{P}_A$ and $\mathcal{P}_B$. We construct the null and alternate hypotheses as the following:

\begin{itemize}
    \item Null hypothesis, $H_0: \hat{p}_A = \hat{p}_B$, i.e., there is no statistically significant difference between the two proportions. Both EM pipelines pass (or fail) with similar probability.
    
    \item Alternative hypothesis, $H_A: \hat{p}_A > \hat{p}_B$ i.e., $ \mathcal{P}_A$ has significantly higher probability of passing than $\mathcal{P}_B$.
\end{itemize}

To accept or reject the Null hypothesis, we set a significance level, $\alpha = 0.05$, and compute the test statistics (Z-score) as follows:
\begin{equation}
    z^*_{(two-sample)} = \frac{\hat{p}_A - \hat{p}_B}{\sqrt{p^* (1 - p^*) (\frac{1}{n_A} + \frac{1}{n_B})}},
\end{equation}
Where the parameters are defined in the following table,
\begingroup
\renewcommand{\arraystretch}{1.5} %
\begin{table}[h]
    \centering
    \begin{tabular}{cl}
    \hline
        Parameter & Description \\
        \hline
      
        $x_A$ & \# of successful experiments in $\mathcal{P}_A$\\
        
        $n_A$ & \# of total experiments in $\mathcal{P}_A$\\
        
        $x_B$ & \# of successful experiments in $\mathcal{P}_B$\\
        
        $n_B$ & \# of total experiments in $\mathcal{P}_B$\\
        
         $p^*$ & $ \frac{(x_A + x_B)}{(n_A + n_B)}$\\
        
        $\hat{p}_A$ & $ \frac{x_A}{n_A}$\\
        
        $\hat{p}_B$ & $ \frac{x_B}{n_B}$\\
    \hline\\
    \end{tabular}
    \caption{Parameters involved in calculating the test statistics.}
    \label{tab:hypothesis_test_params}
\end{table}
\endgroup

\subsubsection*{Confidence interval: two-sample proportions}
After determining a significant difference between two pipelines from hypothesis testing, we compute the $95\%$ CI for the difference between two population proportions as follows:
\begin{equation}
    CI = 100 \times ((\hat{p_A} - \hat{p_B}) \pm z^* \times SE)
\end{equation}

where, standard error $SE = \sqrt{(\hat{p_A}(1-\hat{p_A})/n_A)^2 + (\hat{p_B}(1-\hat{p_B})/n_B)^2}$ and $z^* = 1.96$ for $95\%$ confidence level.

The interval tells us how much more successful the pipeline $\mathcal{P}_A$ is than pipeline $\mathcal{P}_B$. Suppose we get $CI = 8 \pm 2\%$. It will mean that error mitigation with $\mathcal{P}_A$ will generate $6\%$ to $10\%$ more successfully error-mitigated experiments (i.e., $REM < 1$) than $\mathcal{P}_B$ at $95\%$ confidence level.

\section{Experimental Details}\label{sec:em-details}
Having setup the framework for evaluating our experimental results, we now dedicate this section to describe the experimental setup for which the results shall be presented and analyzed. In this work, we evaluate $16$ different pipelines, which are composed of the following error mitigation techniques:  \texttt{ZNE}, \texttt{MEM}~\cite{mem}, \texttt{RC}~\cite{randomized_comp}, \texttt{DD} and mitigation with estimation circuits~\cite{Urbanek2021}. Ref.~\cite{Cai2022} contains a summary of each technique. Table~\ref{tab:em-pipelines} describes the composition of each pipeline considered in this work. All  our experiments were run on two $7$-qubit IBM devices, namely, \texttt{ibm\_lagos} and \texttt{ibm\_perth}, and  physical qubits \texttt{[0, 1, 2, 3]} were used to run circuits on each device.

\begingroup
\renewcommand{\arraystretch}{1.5} %
\begin{table}[h]
    \centering
    \begin{tabular}{cl}
    \hline
        Pipeline & Pipeline composition \\
        \hline
        $\mathcal{P}_1$ ($\mathcal{P}_1^{E}$) & \texttt{ZNE} (\texttt{ZNE}$^{(E)}$)\\
        
        $\mathcal{P}_2$ ($\mathcal{P}_2^{E}$) & \texttt{ZNE} (\texttt{ZNE}$^{(E)}$) + \texttt{MEM}\\
        
        $\mathcal{P}_3$ ($\mathcal{P}_3^{E}$) & \texttt{ZNE} (\texttt{ZNE}$^{(E)}$) + \texttt{DD} (`X-X' sequence) \\
        
        $\mathcal{P}_4$ ($\mathcal{P}_4^{E}$) & \texttt{ZNE} (\texttt{ZNE}$^{(E)}$) + \texttt{DD} (`X-X' sequence) + \texttt{MEM}  \\
        
        $\mathcal{P}_5$ ($\mathcal{P}_5^{E}$) & \texttt{ZNE} (\texttt{ZNE}$^{(E)}$) + \texttt{RC} \\
        
        $\mathcal{P}_6$ ($\mathcal{P}_6^{E}$) & \texttt{ZNE} (\texttt{ZNE}$^{(E)}$) + \texttt{RC} + \texttt{MEM} \\
        
        $\mathcal{P}_7$ ($\mathcal{P}_7^{E}$) & \texttt{ZNE} (\texttt{ZNE}$^{(E)}$) + \texttt{RC} + \texttt{DD} (`X-X' sequence) \\
        
        $\mathcal{P}_8$ ($\mathcal{P}_8^{E}$) & \texttt{ZNE} (\texttt{ZNE}$^{(E)}$)  + \texttt{RC} + \texttt{DD} (`X-X' sequence) + \texttt{MEM} \\
    \hline\\
    \end{tabular}
    \caption{Composition of error mitigation pipelines evaluated in this paper. Each pipeline is either standalone \texttt{ZNE} or other EM methods such as \texttt{MEM}, \texttt{RC}, and \texttt{DD} applied on top of \texttt{ZNE}. We also test the same pipelines where the expectation values fed to \texttt{ZNE} are corrected with noise estimation circuits~\cite{Urbanek2021}. Those pipelines are denoted by an extra subscript $E$ such as $\mathcal{P}_1^{E}$ and mentioned inside parentheses.}
    \label{tab:em-pipelines}
\end{table}
\endgroup

\subsection*{Test Circuits}
    \begin{figure}
        \centering
        \includegraphics[width=6.0in]{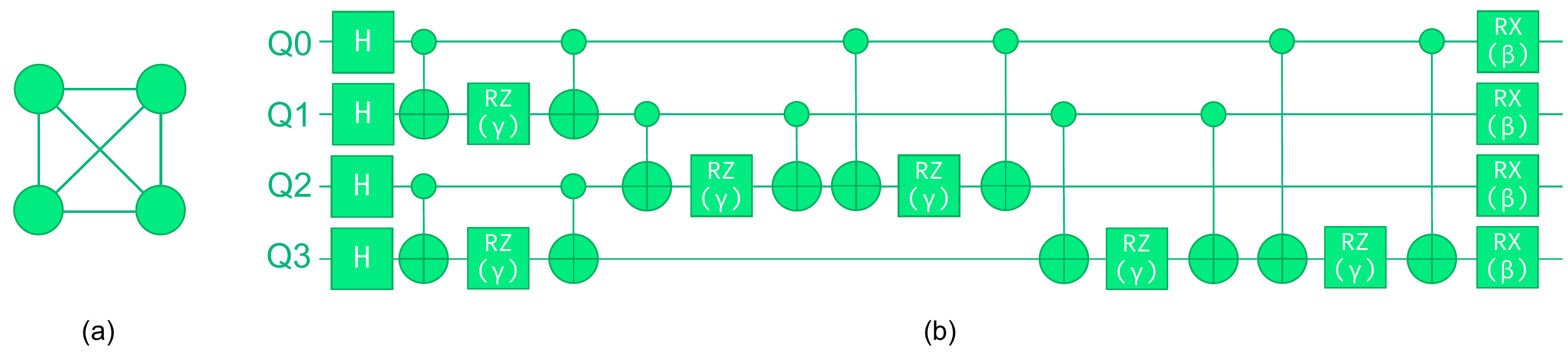}
        \caption{(a) $4$-node all connected graph for MaxCut. (b) Corresponding $4$-qubit QAOA quantum circuit.}
        \label{fig:qaoa-graph-and-circuit}
    \end{figure}
    As the test circuit, we chose the QAOA-MaxCut circuit for all-connected $4$-node graphs with $1$-layer. Each circuit consists of $2$ parameters ($\gamma, \beta$). Figure~\ref{fig:qaoa-graph-and-circuit}a shows the $4$-node all connected graph for the MaxCut problem, and Figure~\ref{fig:qaoa-graph-and-circuit}b shows the corresponding quantum circuit.
    We observed that the success rate of an EM pipeline could have some dependence on the circuit parameters chosen for the test circuit. In order to incorporate this in our analysis,  we selected $10$ pairs of $(\gamma, \beta)$, which cover the ideal expectation value range from $-0.62$ to $2.0$ (excluding the constant term in the operator) for the circuit in Figure~\ref{fig:qaoa-graph-and-circuit}

\subsection*{Zero Noise Extrapolation}
    In zero noise extrapolation (\texttt{ZNE}), expectation values of an algorithm are computed at multiple noise levels. Next, the expectation values are regressed using a choice of fitting functions, such as linear and quadratic, to extrapolate the expectation values in the zero noise limit. We compute expectation values at $3$ noise levels or scale factors in our experiments, \texttt{[1.0, 3.0, 5.0]}. Noise level $= 1.0$ corresponds to the original circuit, whereas to achieve noise levels $3.0$ and $5.0$, the noise in the original circuit has to be amplified. In our experiments, we adopt the digital noise amplification technique~\cite{LaRose2020, Giurgica-Tiron2020a} to amplify the noise. It involves \emph{folding} gates in the original circuit such that the logical function of the circuit remains the same, but the circuit experiences more noise due to elevated gate counts.

\begin{figure}
        \centering
        \includegraphics[width=5.0in]{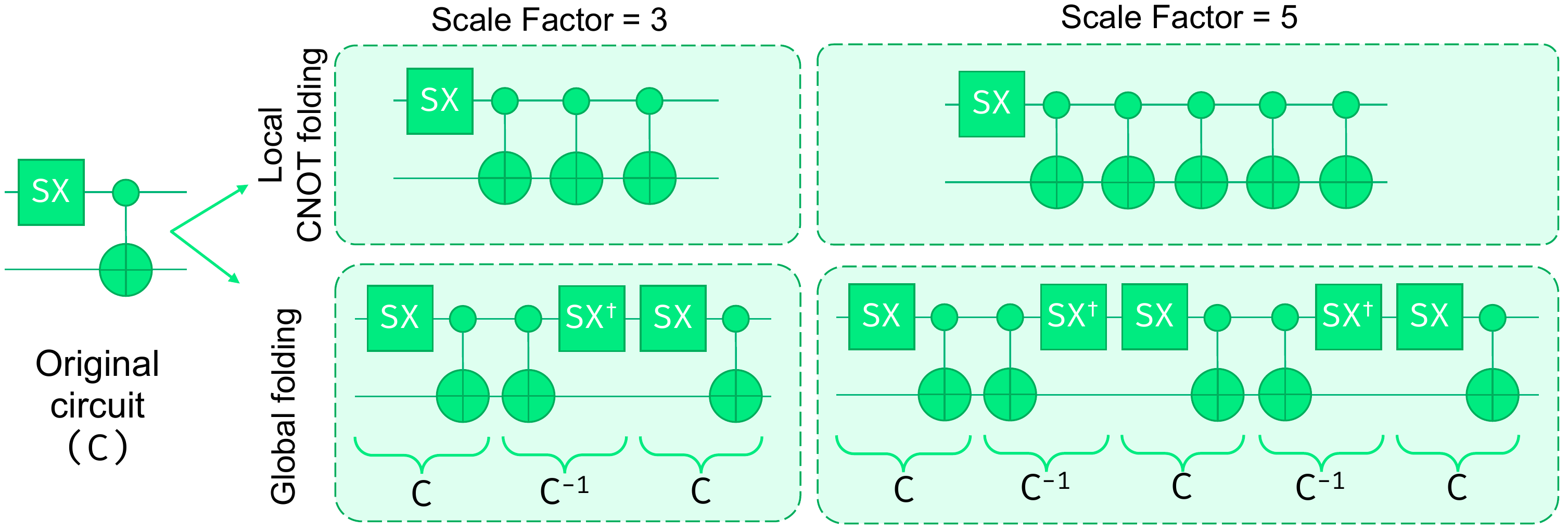}
        \caption{Example of local CNOT and global folding used in zero noise extrapolation experiments.}
        \label{fig:folding}
    \end{figure}

    We employ two types of gate folding techniques, namely, local CNOT folding and global folding. The basic idea of each type of folding is depicted in Figure~\ref{fig:folding}. In the case of local CNOT folding, each CNOT gate is replaced by $3$ or $5$ consecutive CNOT gates to achieve a noise scale factor of $3$ and $5$, respectively. 
    For the global folding, the original circuit ($C$) is appended by its inverse and itself ($C^{-1}C$), once for noise amplification factor $3$ and twice for the factor $5$. 
    
    We use the Qiskit transpiler to get device-executable circuits. Then, circuits corresponding to noise levels $1.0$, $3.0$, and $5.0$ are executed on the hardware, and expectation values are computed. We repeat the experiments $15$ times to get a distribution of expectation values at each noise level. Each of these $15$ data points of expectation values per noise level is computed with $10,000$ shots. Finally, expectation values are fitted using both linear and quadratic regression methods to find the expectation value at the zero noise limit. We use \emph{non-parametric bootstrapping}~\cite{Efron1982} with $10,000$ bootstrap samples to calculate the variance of the zero noise limit result.

\subsection*{Measurement Error Mitigation}
    As the measurement error mitigator, we select the linear algebra-based approach~\cite{mem} available in the Qiskit experiments. It involves running calibration circuits to construct a measurement calibration matrix, $M_{calib}$. The calibration matrix is then inverted and multiplied with the noisy counts to get the measurement error mitigated counts, $Count_{mitigated} = M_{calib}^{-1}Count_{noisy}$. For $n$ measured qubits, the process involves running $2^n$ calibration circuits ($2^4=16$ calibration circuits for $4$-qubit experiments). We run each measurement calibration circuit with $10,000$ shots.
    
\subsection*{Randomized Compilation}
\begin{figure}
        \centering
        \includegraphics[width=6in]{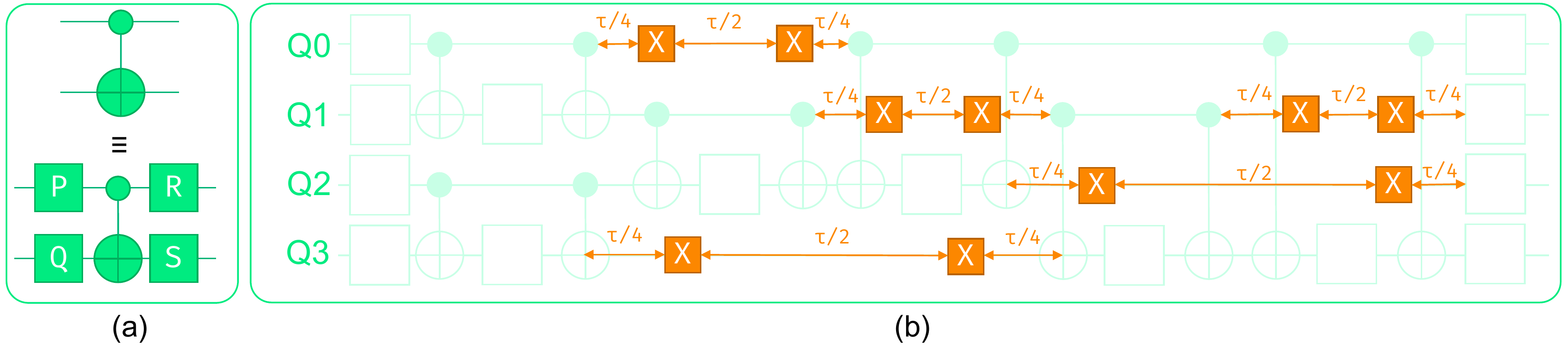}
        \caption{(a) Dressing of CNOT gate for randomized compilation (\texttt{RC}). Two Pauli gates each are added before and after a CNOT gate so that it logically remains a CNOT. (b) Dynamical Decoupling. Idle time on a qubit between two gates is dressed with the Idle ($\tau/4$) - X - Idle ($\tau/2$) - X - Idle ($\tau/4$) sequence. (The length of the arrows denoting the $\tau/2$ and $\tau/4$ times are not to scale.)} 
        \label{fig:rc-and-dd}
    \end{figure}

    Randomized compilation involves generating multiple random duplicates of an original circuit and aggregating the counts of duplicates to get a noise-mitigated count.
    In this work, we follow the \texttt{RC} implementation used in~\cite{Urbanek2021}. It involves dressing CNOT gates in a circuit with Pauli gates as shown in Figure~\ref{fig:rc-and-dd}a, such that it logically remains a CNOT. Here, two Pauli gates (\texttt{P, Q}) are added before the CNOT gate, and two (\texttt{R, S}) are added after the CNOT. These \texttt{P, Q, R,} and \texttt{S} gates are chosen independently and randomly for each CNOT gate in the circuit from Table~\ref{tab:rc-dressing}.  
    By dressing CNOT gates as above, we construct $50$ random duplicates per noise-scaled circuit and run each random duplicate with $200$ shots so that we have $50 \times 200 = 10,000$ shots in aggregate.
    
    \begingroup
    \setlength{\tabcolsep}{2pt}
    \renewcommand{\arraystretch}{1.5}

    \begin{table}[]
    \centering
        \begin{tabular}{ccccccccccccccccccc}
        \hline
        $P$ & $Q$ & $R$ & $S$ & \hspace{5mm} & $P$ & $Q$ & $R$ & $S$ & \hspace{5mm} & $P$ & $Q$ & $R$ & $S$ & \hspace{5mm} & $P$ & $Q$ & $R$ & $S$ \\
        \hline
        $I$ & $I$ & $I$ & $I$ &    & $Y$ & $I$ & $Y$ & $X$ &    & $X$ & $I$ & $X$ & $X$ &    & $Z$ & $I$ & $Z$ & $I$ \\
        $I$ & $X$ & $I$ & $X$ &    & $Y$ & $X$ & $Y$ & $I$ &    & $X$ & $X$ & $X$ & $I$ &    & $Z$ & $X$ & $Z$ & $X$ \\
        $I$ & $Y$ & $Z$ & $Y$ &    & $Y$ & $Y$ & $X$ & $Z$ &    & $X$ & $Y$ & $Y$ & $Z$ &    & $Z$ & $Y$ & $I$ & $Y$ \\
        $I$ & $Z$ & $Z$ & $Z$ &    & $Y$ & $Z$ & $X$ & $Y$ &    & $X$ & $Z$ & $Y$ & $Y$ &    & $Z$ & $Z$ & $I$ & $Z$\\
        \hline
        \end{tabular}
        \caption{Choice of randomization gates for \texttt{RC}.}
        \label{tab:rc-dressing}
    \end{table}

    \endgroup
    
\subsection*{Dynamical Decoupling}
    In the case of dynamical decoupling, a sequence of pulses is applied to the system with the purpose of decoupling it from the effects of the environment, usually to qubits that are temporarily idle. Ideally, we would like to apply infinitely many fast pulses, but we choose to work at the digital level of the control stack by applying the more coarse-grained circuit gates instead. For dynamical decoupling, we have a number of options for the gate sequence to apply, but we found that the simplest sequence, `X-X', worked best in the case of our experiments. Figure~\ref{fig:rc-and-dd}b visually shows the idea of \texttt{DD}. Any qubit idle time is dressed with the following sequence of two `X' gates: $\tau/4$ (idle) -- X -- $\tau/2$ (idle) -- X -- $\tau/4$ (idle).
    
\subsection*{Estimation Circuits}
    Estimation circuits~\cite{Urbanek2021} are constructed by removing all single-qubit gates and keeping only the two-qubit CNOT gates in the circuit. Mitigation with estimation circuits involves computing a noise parameter $p$, where $1 - p = \langle \sigma_z^{\otimes n} \rangle$. 
    From hardware experiments, we observed that the noise might be too high for deeper circuits (e.g., in $4$-qubit circuits with a noise scale factor of $5$), such that $\langle \sigma_z^{\otimes n} \rangle$ tends to 0. Correcting the noisy expectation $\langle E_{noisy} \rangle$ value by dividing it by $1-p$ ($= \langle \sigma_z^{\otimes n} \rangle \rightarrow 0$) overshoots the corrected expectation values. Thus, we modify the noise parameter $p$ such that 
    \begin{equation}
      \label{eq:modified_noise_parameter}
      1-p =\frac{\# b_{000\dots 00}}{N}, 
    \end{equation}
    where $\# b_{000\dots 00}$ is the number of all-zero bitstrings and $N$ is the total number of shots. Mitigation with a modified definition of the noise parameter $p$ provided better results in our experiments.

\section{Resource vs. Mitigation Efficiency Trade Off Considerations}\label{sec:metric}
Having discussed how to judge different error mitigation techniques, we now deal with another equally important problem: how do we quantify the resource requirement for accuracy improvement? Some error mitigation methods require an increase in depth, while others require an increase in the number of circuits to run. Therefore, it is necessary to devise a figure of merit that captures these different overhead concerns. Also how do we account for the \emph{complexity} of an error mitigation pipeline? In this section, we introduce a classical information theory-inspired entropic figure of merit to capture the overhead and complexity of error mitigation pipelines and name it resource, $R$. Finally, we combine $R$ with two measures of error mitigation efficiency and introduce a single resource normalized metric for the overall quality of an error mitigation pipeline.

\subsection{An Entropic Figure of Merit for Resource}
Let label $R$ be figure of merit that includes both the overhead and the complexity. An obvious choice for $R$ is to count the number of \emph{shots} or \emph{samples} taken on the quantum computer. While this is a good first-order approximation, it only partially captures the situation. For example, if mitigation method $A$ requires $1$ circuit with $10,000$ shots and mitigation method $B$ requires $10$ circuits with $1,000$ shots each, both require the same number of shots on the quantum computer. However, method $B$ will be more challenging to run as the program must be modified during the experiment. This will introduce additional overhead in the classical control/support hardware and increase the challenge of scheduling circuits (jobs) on a cloud computing service. There are two options on hand, one could either develop models of run-time for each hardware provider or invent some proxy for the complexity of running experiments. This proxy should be calculable independent of the hardware provider and easily interpretable. We arrive at a proxy using the following reasoning: what increases the complexity of an algorithm from the experimental point of view are the number of \textit{distinct} circuits one needs to run. We therefore consider an algorithms that needs to run fewer distinct circuits to be simpler than one that needs to run more distinct circuits. From this point of view
 the concept of \emph{entropy} is a natural measure to consider. We define the resource metric $R$ by the following equation:
\begin{equation}\label{eq:resource}
    R = T(1+S)
\end{equation}
where $T$ is, for now, the total number of \emph{shots} or \emph{samples} taken on the quantum computer, and $S$ is the entropy which is defined as: 

\begin{equation}
    S = -\sum_{i} p_{C_i} \ln(p_{C_i})
    \label{eq:S}
\end{equation}
where $p_{C_i}$ is the probability of the distinct circuit $C_i$ where each $C_i$ can be thought of as a symbol for an error mitigation $\mathcal{A}$ , i.e.,  $\mathcal{A}= \{ C_1, C_2, \dots C_N \}$.
The question then becomes, what probability $p_{C_i}$ should we attach to these circuits?
At first pass, $p_{C_i} = \frac{N_{C_i}}{N} $,  where $N_{C_i}$ is the number of shots run for circuit $C_i$ and $N= \sum_i N_{C_i}$ is the total number of shots needed for algorithm $\mathcal{A}$. 

However, not all circuits in an algorithm will necessarily have the same number of gates or depth or duration. This should also be accounted for. For each circuit $C_i$, let $D_{C_i}$ be its duration. 
We then need an updated measure for quantum computer usage. We can use the duration-weighted total number of shots to estimate the time spent running on the quantum hardware.
\begin{equation}
     \label{eq:Time}
    T = \sum_i N_{C_i} D_{C_i}
\end{equation}

Finally, different circuits required to implement the algorithm $\mathcal{A}$ may need a different number of qubits. If the circuit is dense enough and parallelization of all gates is impossible then in some sense a circuit with more qubits is harder to implement than one with fewer qubits. Direct comparison of circuits with different number of qubits is not clear but we can modify  (\ref{eq:Time}) to a qubit number weighted average as follows:

\begin{equation}
    T = \sum_i N_{C_i} D_{C_i} Q^{norm}_{C_i}
\end{equation}
where, $Q^{norm}_{C_i} = Q_{C_i} / Q_{max}$ is the qubit count of circuit $C_i$ normalized by maximum qubit count ($Q_{max}$) among circuits in $\mathcal{A}$.

We can thus define the probability of a circuit be 
\begin{equation}
    p_{C_i} = \frac{ N_{C_i} D_{C_i} Q^{norm}_{C_i}}{T}
\end{equation}

\subsection{Combining Mitigation Efficiency and Resource}
Now that we have defined an entropic figure of metric that takes into account the total number of circuits, the number of shots for each circuit, and the duration of each circuit, we include the error mitigation efficiency achieved by the algorithm. 
We use $REM$ to measure the mitigation efficiency as it quantifies to what extent an error mitigation method can push the mitigated expectation value to the ideal value—the lower the $REM$ value, the better the mitigation. We compute the $95\%$ confidence interval of the median $REM$ of a pipeline and take the upper interval in our calculations to be conservative. Next, we argue that only accounting for $REM$ values may not tell us the complete picture. A pipeline may have overall low $REM$ but fail to mitigate errors more frequently than others. Thus, we need to consider the proportion of \texttt{SUCCESS} (i.e., how often a pipeline results in REM $< 1$) of the pipeline along with median $REM$. We compute the $95\%$ confidence interval of the proportion of \texttt{SUCCESS} and, this time, take the lower interval to be conservative. We name the lower interval as \emph{pipeline success rate} ($PSR$). The higher the $PSR$, the better the pipeline. 

Finally, the overall quality of an error mitigation pipeline depends on three factors: resource ($R$), $REM$, and $PSR$. Among these three factors, we want $R$ and median $REM$ ($\epsilon$) to be lower and $PSR$ to be higher. Thus, we can combine them to compute a single metric for the overall quality of mitigation, $M$, as follows:

\begin{equation}
    M = \frac{PSR(\%)}{\epsilon \times R}
    \label{eq:mitigationMetric}
\end{equation}

Eq.~\ref{eq:mitigationMetric} can be tweaked in different ways, such as one can take weighted versions of each factor to prioritize one over the other. Instead of $REM$, $\epsilon$ can represent different metrics depending on the problem. While computing $R$, $D$ can be taken as more generic \emph{depth} (or, \emph{weighted-depth}) of a circuit instead of the device-dependent raw duration. However, in this paper, we take the non-weighted version of the parameters with $\epsilon$ as $REM$ and $D$ as the duration (in seconds) of a quantum circuit.

\section{Results and Analysis}\label{sec:results}
\subsection{Data Preparation and Analysis}
 As per Eq.~\ref{eq:rem}, three parameters are needed to compute an $REM$ value, namely $\langle E \rangle_{ideal}$, $\langle E \rangle_{\lambda=0}$ (expectation value in the zero noise limit, i.e., the mitigated expectation value $\langle E \rangle_{mitigated}$), and $\langle E \rangle_{\lambda=1}$ (expectation value with no scaling of gates or circuits, i.e., the noisy expectation value $\langle E \rangle_{noisy}$). 

 The mean ($\mu_{\lambda=1}$) and standard deviation ($\sigma_{\lambda=1}$) of $\langle E \rangle_{\lambda=1}$ are computed from running $15$ experimental runs of original circuit at noise level $1$ for a specific set of parameters.
 Using \emph{non-parametric bootstrapping}~\cite{Efron1982}, we approximated the mean ($\mu_{\lambda=0}$) and standard deviation ($\sigma_{\lambda=0}$) of the regression parameter, $\langle E \rangle_{\lambda=0}$. 
 As the statistical hypothesis testing framework described in Section~\ref{sec:hypothesis-testing} works on binarized $REM$ values, we follow the procedure in Algorithm~\ref{alg:data_prep} to produce data for the hypothesis testing framework.

    \begin{algorithm}
    \caption{Characterize Success Rate for an EM pipeline}\label{alg:data_prep}
    \hspace*{\algorithmicindent} \textbf{Input}:  $\{\gamma^{(i)}, \beta^{(i)} \}_{i=1}^{i=10}$, \{$\mu_{\lambda=0}^{(i)}, \sigma_{\lambda=0}^{(i)}\}_{i=1}^{i=10},\{ \mu_{\lambda=1}^{(i)}, \sigma_{\lambda=1}^{(i)} \}_{i=1}^{i=10} $ \\
    \hspace*{\algorithmicindent} \textbf{Output}: Success and Failure Counts for an EM pipeline
    \begin{algorithmic}[1]
    \Procedure{Data Preparation}{}
    \State $\texttt{success} \gets 0$
    \State $\texttt{failure} \gets 0$

     \For{$i=1$ to $i=10$}
    \State Generate 1000 samples for $\langle E \rangle_{\lambda=0}^{(i)} $ from      $\mathcal{N}(\mu_{\lambda=0}^{(i)},\sigma_{\lambda=0}^{(i)})$
    \State Generate 1000 samples for $\langle E \rangle_{\lambda=1}^{(i)} $ from $\mathcal{N}(\mu_{\lambda=1}^{(i)},\sigma_{\lambda=1}^{(i)})$
     \State Plug values into Eq.~\ref{eq:rem}  to get 1000 REM Values
     \For{$j=1$ to $j=1000$}
         \If{$REM < 1$}
             \State $\texttt{success} \gets \texttt{success} + 1 $
        \ElsIf{$REM \geq  1$}
              \State $\texttt{failure} \gets \texttt{failure} + 1 $
       \EndIf
      \EndFor
      \EndFor
     \State Return \texttt{Success}, \texttt{Failure}
    \EndProcedure
    \end{algorithmic}
    \end{algorithm}

 The statistical tests are applied on the experimental data collected from hardware experiments on the \texttt{ibm\_lagos} and \texttt{ibm\_perth}. We ran $91,880$ circuits on each device per folding type to collect the hardware data. 
\begin{figure}[b]
    \centering
    \includegraphics[width=6.5in]{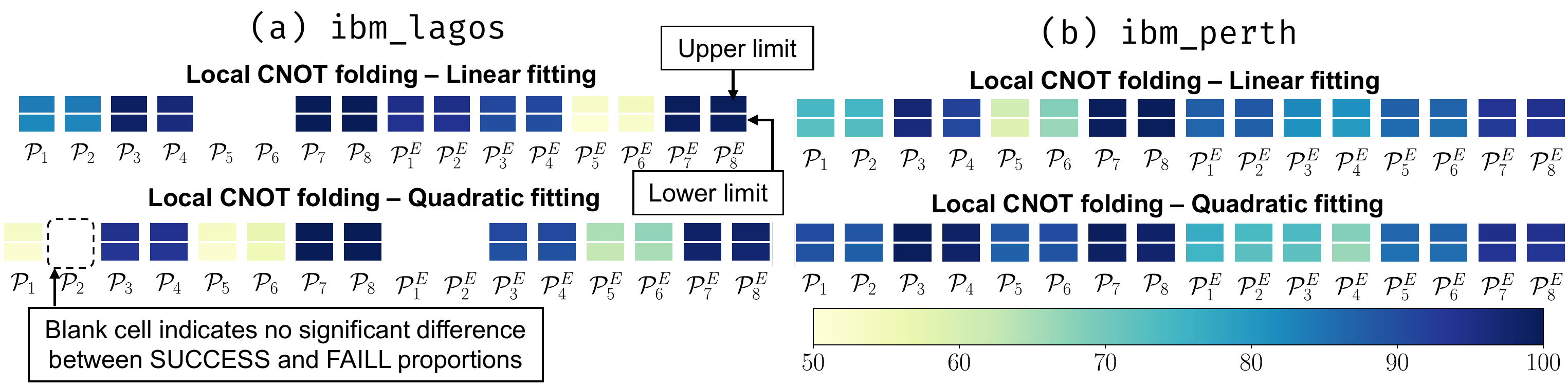}
    \caption{Confidence intervals of proportions of successful experiments for two devices, \texttt{ibm\_lagos} and \texttt{ibm\_perth}, and both linear and quadratic fitting. Bluer cells indicate a higher proportion of success.  $\mathcal{P}_3$, $\mathcal{P}_4$, $\mathcal{P}_7$, $\mathcal{P}_8$, $\mathcal{P}_7^{E}$, and $\mathcal{P}_8^{E}$ resulted in bluest intervals across devices and fitting types in general. Blank cells indicate that the proportion of \texttt{SUCCESS} is not significantly higher than the proportion of \texttt{FAIL} for a pipeline. This observation emphasizes the importance of assigning statistical confidence to the mitigation capability of EM pipelines. A user may achieve mitigation with blank pipelines at times, but the performance will be inconsistent, which is paramount.}
    \label{fig:one_prop_conf_int}
\end{figure}

When evaluating the EM pipelines individually (Figure~\ref{fig:one_prop_conf_int}), we see  that $\mathcal{P}_3$, $\mathcal{P}_4$, $\mathcal{P}_7$, $\mathcal{P}_8$, $\mathcal{P}_7^E$, and $\mathcal{P}_8^E$ are the best for both devices and for both fitting types. Each of these pipelines has dynamical decoupling (\texttt{DD}) in common, which hints towards an essential role of \texttt{DD} on these devices. Already we are beginning to see in our toy experiments how trade off decisions can be important. Randomized Compiling can increase the number of circuits one needs to run by an order of magnitude or 2  but if it is the case the \texttt{DD} is what is playing the essential role in mitigation and producing results that are close to more complicated pipeline, there is some room left for resource considerations.

An interesting observation from the \texttt{ibm\_lagos} device is that a few pipelines on this device are failing to generate significantly higher proportion of \texttt{SUCCESS} than \texttt{FAIL}, and those cells are left blank in the confidence interval plot in Figure~\ref{fig:one_prop_conf_int}a. This does not mean that the pipelines \emph{always} fail to mitigate error, instead the pipelines may mitigate errors in some cases, but the probability is not distinguishable from a $50$-$50$ draw or is in fact biased towards increasing the error within our confidence.
This observation highlights the importance of attaching a statistically-supported confidence to pipelines. We also consider the probability of type II errors, i.e., if we fail to reject the null hypothesis and conclude that an EM pipeline does not mitigate error more frequently when in reality it does. Due to large sample size ($10,000 REM$ values), the probability of type II error in our cases was practically zero.

\begin{figure}
    \centering
    \includegraphics[width=6in]{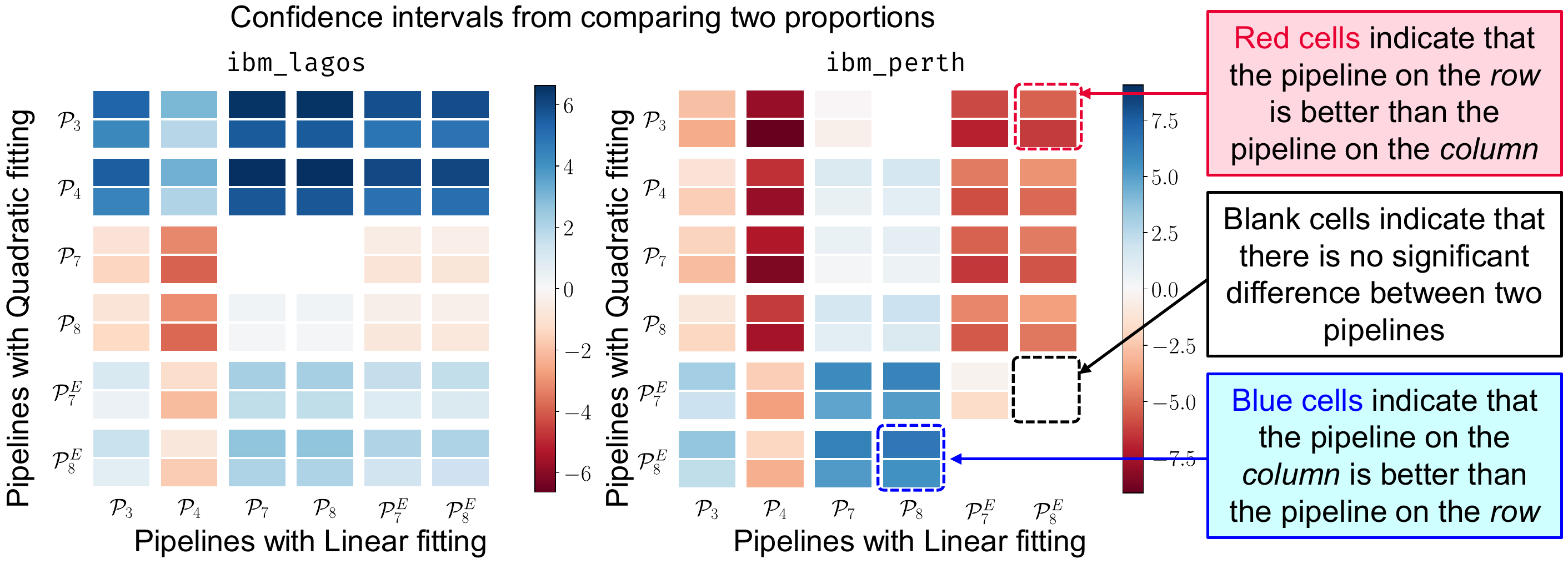}
    \caption{$95\%$ confidence intervals for comparison between two pipelines. A more positive confidence interval (blue) indicates that the pipeline on the column (X-axis) has a higher proportion of \texttt{SUCCESS} than the pipeline on the row. In contrast, a negative (red) interval means the opposite. Blank cells indicate no statistically significant difference between the two compared pipelines.}
 
    \label{fig:conf-int}
\end{figure}

Next, we compare the linear and quadratic fitting variants of the $6$ best-performing pipelines, i.e., $\mathcal{P}_3$, $\mathcal{P}_4$, $\mathcal{P}_7$, $\mathcal{P}_8$, $\mathcal{P}_7^E$, and $\mathcal{P}_8^E$ using the two-sample test of proportions. The confidence intervals are plotted in Figure~\ref{fig:conf-int}. A blue cell indicates that the pipeline on the column (linear fitting) has a better proportion of \texttt{SUCCESS} than the pipeline on the row (quadratic fitting), and a red cell specifies the opposite. The diagonal elements of the heatmaps compare the linear and quadratic variants of the same pipeline. 
On \texttt{ibm\_lagos}, 
linear variants of $\mathcal{P}_7$ and $\mathcal{P}_8$ perform better than most quadratic variants in general, except $\mathcal{P}_7$ (quadratic). $\mathcal{P}_7$ (quadratic) does not have a statistically significant difference from either  $\mathcal{P}_7$ (linear) or $\mathcal{P}_8$ (linear). On the other hand, quadratic variants of $\mathcal{P}_3$ and $\mathcal{P}_4$ perform worse than all linear variants on \texttt{ibm\_lagos}.

On \texttt{ibm\_perth}, the trend is different. Inspecting the diagonal elements provides a mixed scenario. For instance, quadratic versions of $\mathcal{P}_3$, $\mathcal{P}_4$, and $\mathcal{P}_7^E$ have better success proportions than their respective linear versions. In contrast, $\mathcal{P}_7$ and $\mathcal{P}_8$ quadratic versions are marginally worse than their linear counterparts. We observe no significant difference between linear and quadratic fittings in the case of $\mathcal{P}_8^E$. $\mathcal{P}_3$ (quadratic) and $\mathcal{P}_8^E$ (linear) have the edge over other pipelines except between themselves as $\mathcal{P}_3$ (quadratic) has all red cells in the row whereas $\mathcal{P}_8^E$ (linear) has all blue cells on the column. \textit{The results from two different devices indicate that the choice of fitting function is device dependent.}

\subsection{Resources vs. Mitigation Efficiency}
In this section, we analyze the resource vs. (mitigation) efficiency of different pipelines using the metrics proposed in Sec.~\ref{sec:metric}. As a first pass, the metric $M$ provides a user with a single metric to compare and choose a pipeline for error mitigation. The user may choose the pipeline with highest $M$. 

\begingroup
\renewcommand{\arraystretch}{1.5}

\begin{table}[]
\centering
\begin{tabular}{ccccccccc}
\hline
Device                     & Pipeline & R       & REM (lin)    & REM (quad)   & PSR (lin)   & PSR (quad)   & M (lin)       & M (quad)       \\
\hline
\multirow{4}{*}{\texttt{ibm\_lagos}} & $\mathcal{P}_3$       & $2.8494$  & $0.5397$ & $0.3126$ & $0.9858$ & $0.9352$ & $64.1009$ & $104.9981$ \\
                           & $\mathcal{P}_4$       & $3.8006$  & $0.5097$ & $0.2435$ & $0.9609$ & $0.9339$ & $49.6020$ & $100.9129$ \\
                           & $\mathcal{P}_8$       & $9.9802$  & $0.2815$ & $0.0992$ & $1.0000$ & $0.9981$ & $35.5945$ & $100.8167$ \\
                           & $\mathcal{P}_7^E$      & $20.1687$ & $0.1170$ & $0.1240$ & $0.9915$ & $0.9780$ & $42.0164$ & $39.1058$  \\
                           \hline
                           % \hline
\multirow{3}{*}{\texttt{ibm\_perth}} & $\mathcal{P}_3$       & $2.3061$  & $0.6899$ & $0.4267$ & $0.9623$ & $0.9948$ & $60.4875$ & $101.0925$ \\
                           & $\mathcal{P}_4$       & $3.1189$  & $0.5950$ & $0.3007$ & $0.9055$ & $0.9811$ & $48.7956$ & $104.6124$ \\
                           & $\mathcal{P}_7^E$      & $16.5019$ & $0.1578$ & $0.1625$ & $0.9258$ & $0.9363$ & $35.5540$ & $34.9154$ \\
                           \hline
\end{tabular}
\caption{Resource usage and mitigation efficiency values for partial set of pipelines. This table exemplifies the resource-efficiency trade-offs in error mitigation. One can compose more complex EM pipeline by combining various EM methods and achieve better mitigation. However, this improved mitigation may come at the cost of more resources. This contrasting dynamics is reflected on the $M$ values of the pipelines. For example, $\mathcal{P}_7^E$ on both devices high $PSR$ and low $REM$ values. However, due to significant resource needs, the pipelines have the lowest $M$ values.}
\label{tab:clipped}
\end{table}

\endgroup

Consider Table~\ref{tab:clipped} which tabulates the resources and accuracy for a partial set of pipelines (full set of values are tabulated in Table~\ref{tab:lagos-local} for \texttt{ibm\_lagos} and in Table~\ref{tab:perth-local} for \texttt{ibm\_perth}). $\mathcal{P}_3$, $\mathcal{P}_4$, and $\mathcal{P}_8$ are the top three pipelines in terms of $M$ values for \texttt{ibm\_lagos}. While $\mathcal{P}_8$ has better $REM$ and $PSR$ values, its $M$ is slightly inferior to $\mathcal{P}_3$ due to higher resource requirements. An extreme case of the trade-offs between accuracy and resources required can be seen for  $\mathcal{P}_7^E$ which has the lowest median $REM$, however mitigation with estimation circuits almost doubles the required number circuit runs leading to a significantly higher resource value. This leads to a low $M$ value. We observe similar trend in the \texttt{ibm\_perth} as well with $\mathcal{P}_3$ and $\mathcal{P}_4$ having the top two $M$ values and $\mathcal{P}_7^E$ getting penalized for resource consumption.

The metric $M$ equips a user with a simple  tool to assess an EM pipeline. In addition to the tables, we introduce the concept of \emph{resource-efficiency space} which visually shows the trade-offs between resources and  mitigation efficiency. The resource-efficiency space is a scatter plot where each marker is placed at a point with the X-coordinate as the inverse of $REM$ and the Y-coordinate as the $PSR$. The marker size corresponds to the resource ($R$), i.e., the more prominent marker, the higher the resource requirement of the pipeline. We expect a pipeline with a lower resource requirement (i.e., smaller markers) to have a higher $PSR$ and a lower $REM$. Thus, the upper-right corner is the favorable spot in the resource-efficiency space.

\begin{figure}[t]
    \centering
    \includegraphics[width=6in]{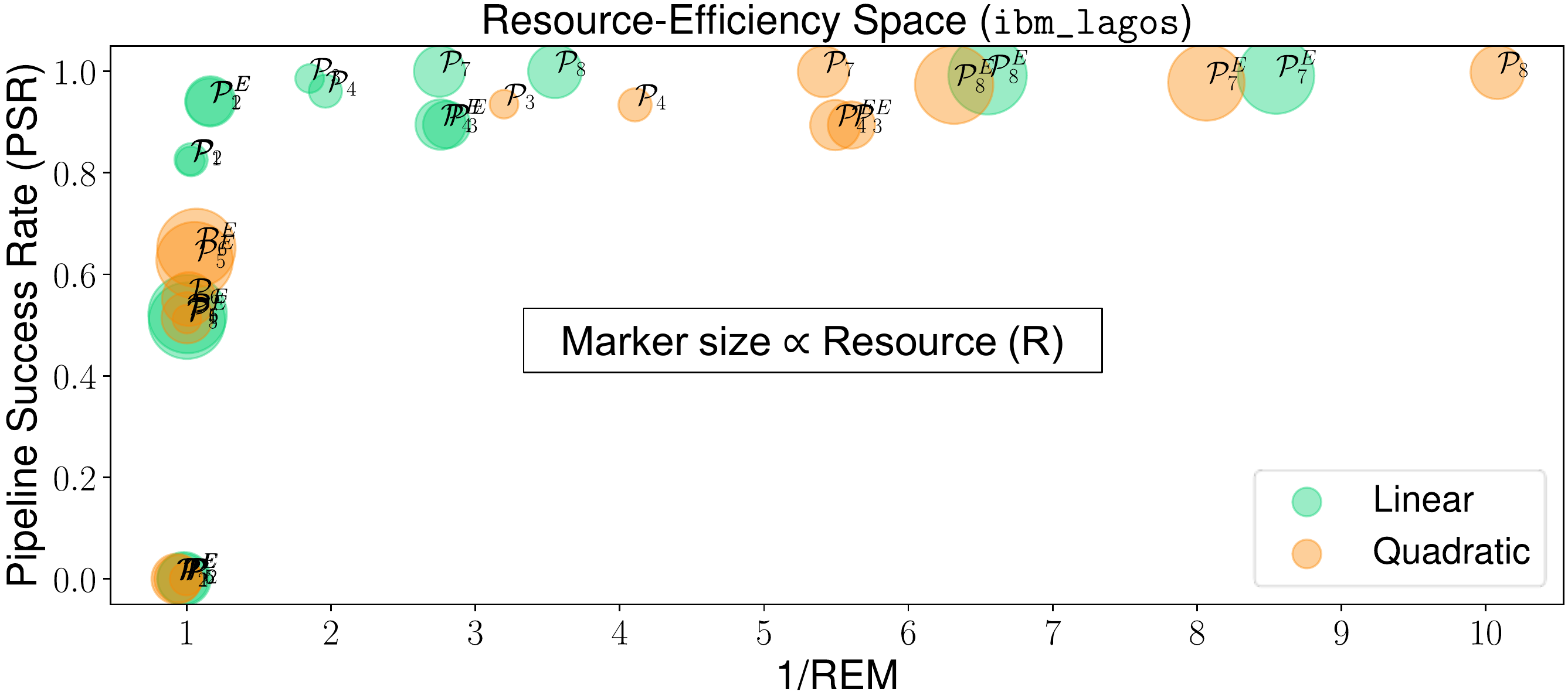}
    \caption{Resource-Efficiency space for experiments on \texttt{ibm\_lagos} with local folding \texttt{ZNE}. $\mathcal{P}_8$ with quadratic fitting stands out among other pipelines in terms of mitigation efficiency as it has a high $PSR$ and the lowest $REM$. } 
    \label{fig:resource-vs-accuracy-lagos-local}
\end{figure}

\begin{figure}[t]
    \centering
    \includegraphics[width=6in]{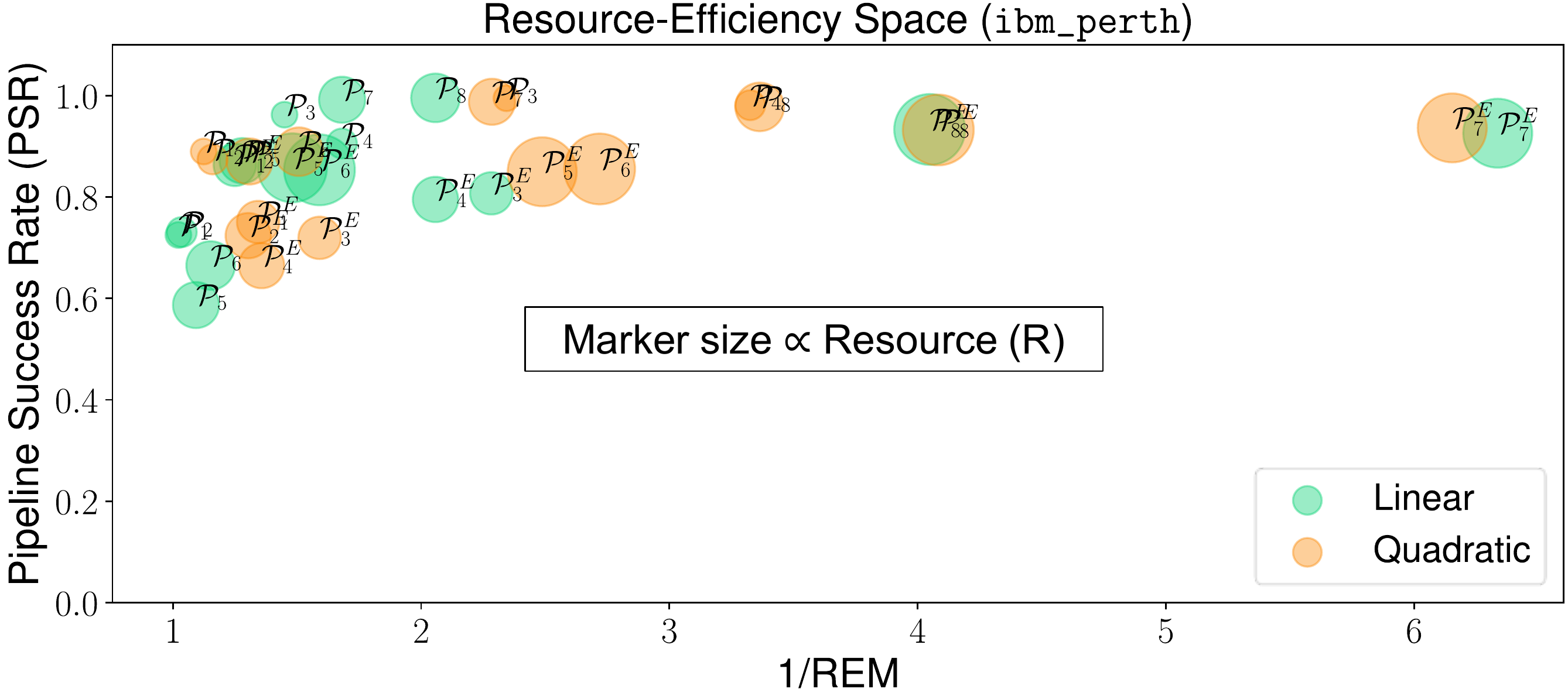}
    \caption{Resource-Efficiency space for experiments on \texttt{ibm\_perth} with local folding \texttt{ZNE}. }
    \label{fig:resource-vs-accuracy-perth-local}
\end{figure}

Figure~\ref{fig:resource-vs-accuracy-lagos-local} shows the resource-efficiency space for \texttt{ibm\_lagos} for both linear and quadratic fitting. 
It visually displays the insights from clipped Table~\ref{tab:clipped}. For instance, $\mathcal{P}_8$ with quadratic fitting stands out among other pipelines in terms of mitigation quality on \texttt{ibm\_lagos}. It has high $PSR$ and $1/REM$ with modest resource expenditure. Thus, if mitigation quality is the priority for users, they should choose $\mathcal{P}_8$ (quadratic). On the other hand, if users have resource constraints, then they may opt for pipelines with smaller resources such as $\mathcal{P}_3$ (quadratic) and $\mathcal{P}_4$ (quadratic). These pipelines have lower resource needs (smaller markers) with reasonable mitigation qualities ($PSR$: $93.52\%$ and $93.39\%$ and $REM$: $0.3126$ and $0.2435$, respectively). 

The plots also tell us that expending more resources does not guarantee monotonically improved mitigation efficiency. For example, $\mathcal{P}_5^E$ and $\mathcal{P}_6^E$ have two of the highest resource consumption while providing only modest mitigation. Besides, most pipelines are located in the upper left corner of the space, which means most pipelines can mitigate errors frequently ($PSR \uparrow$); however, the extent of mitigation is small ($REM \uparrow$).

On \texttt{ibm\_perth}, we again observe a crowded upper-left corner in the plot. However, the choice of pipeline on this device is different from \texttt{ibm\_lagos}. If the user prioritizes the mitigation quality, then $\mathcal{P}_7^E$ (both linear and quadratic variants) is the pipeline of choice as it is located in the upper right corner of the plot ($PSR \uparrow$, $REM \downarrow$). If less resource is the priority, then $\mathcal{P}_4$ with quadratic fitting is a decent compromise between quality and resource. Nonetheless, $\mathcal{P}_7$ and $\mathcal{P}_8$, both linear and quadratic variants, top the chart with the highest $PSR$s, i.e., these pipelines guarantee a statistically higher chance of mitigating errors. This trend is common in both \texttt{ibm\_lagos} and \texttt{ibm\_perth} for both fitting types. Lastly, one could note that the X-axis ($1/REM$) of the resource-efficiency space is longer for \texttt{ibm\_lagos} than the X-axis for \texttt{ibm\_perth}, which potentially indicates that the \texttt{ibm\_lagos} may be a better device than \texttt{ibm\_perth}.

Finally, by analyzing the resource-quality space, we can gather the following insights:
\begin{itemize}
    \item The choice of the pipeline with the best quality error mitigation and fitting type is device dependent. Our proposed statistical analysis framework can aid quantum cloud users in selecting the best-performing pipeline.

    \item Adopting a full-fledged pipeline with \texttt{ZNE}, \texttt{MEM}, \texttt{RC}, and \texttt{DD} is a statistically safe choice. 

    \item While there is a resource vs. quality trade-offs among pipelines, adding more resources does not always guarantee a better error mitigation quality.

    \item The choice of an error mitigation pipeline (and +device) depends on the user's priority between resource and mitigation quality.
\end{itemize}

\section{Conclusion}\label{sec:conclusion}
 `\emph{How good is my quantum error mitigation?}' is an open question in the community. In this paper, we formalized the answer to this question with statistical hypothesis testing and introduced a more inclusive measure for resource and mitigation efficiency of any EM method. Our work will enable researchers to evaluate their quantum error mitigation techniques formally.
We demonstrated this framework by experimentally evaluating the performance of $16$ quantum error mitigation pipelines composed of one or more atomic techniques such as zero noise extrapolation, measurement error mitigation, randomized compilation, and dynamical decoupling.   We introduced the use of the one-sample test of proportions to determine if the proportion of \texttt{SUCCESS} (i.e., $REM < 1$) of a pipeline is significantly higher than $0.5$ and the use of the two-sample test of proportions to evaluate if one pipeline has a significantly higher proportion of \texttt{SUCCESS} than the other.

 As error mitigation generally requires more circuits, shots, and maybe more qubits, we introduced an entropic figure of merit  that succinctly incorporates the number of circuits, shots, and qubit count. Finally, we combined the measure of pipeline success ($PSR$), amount of mitigation ($REM$), and resource consumption ($R$) in a single metric, $M$, for overall mitigation. Although the metric $M$ is aligned with the shots normalized $REM$ as in~\cite{russo2022}, the accounting for the number of circuits and qubit usage, in addition to shots, makes it a complete metric. While a single metric for overall mitigation is easy to follow, it may mask the interplay among factors. Thus, we introduce the concept of \emph{resource-efficiency space}, which visualizes the landscape of mitigation efficiency ($PSR$ and $REM$) and resource ($R$) trade-offs.

The evaluation frameworks and metrics we proposed are extensible to different hardware, test circuits, and error mitigation methods. For instance, we can bring probabilistic error cancellation and virtual state distillation and compose new EM pipelines, which can be evaluated similarly using the hypothesis testing framework. Besides, in our experiments, we adopted the digital version of \texttt{ZNE} while a pulse-level \texttt{ZNE}~\cite{Kandala2019a} is available in the literature. It remains an exciting prospect to evaluate pipelines with pulse-level \texttt{ZNE}. Pulse-level ZNE may enable shorter duration circuits and hence, a lower resource $R$, owing to more granular control over noise amplification by pulse stretching. It can also enable mitigating errors in deeper circuits. However, pulse re-calibration may be required for the stretched pulses, which consumes additional resources. Thus, understanding the compromises between gains from shorter duration circuits and loss from pulse re-calibration remains an important open question, along with the comparison of mitigation efficiencies of digital and pulse-level \texttt{ZNE}. 

Another direction of analysis can be testing the same pipeline but with different resource expenditures. For example, pipeline $\mathcal{P}_5$ (\texttt{ZNE} + \texttt{RC}) can be run with more shots per random duplicate and/or with more random duplicates. The statistical tests and resource-efficiency analysis can reveal if there is an efficiency gain from more resources and what is a sweet-spot for resource vs. efficiency of a pipeline.

All things considered, we propose a flexible and formal statistical testing framework and metrics for evaluating error mitigation techniques in this paper. Using these techniques, researchers can perform a wide range of analyses and better assess their mitigation methods.

\begingroup
\renewcommand{\arraystretch}{1.5} %

\begin{table}[]
\centering
\begin{tabular}{cccccccccc}
\hline
Pipeline          & T      & S      & R       & REM    & REM    & PSR    & PSR    & M       & M        \\
                  &        &        &         & (lin)  & (quad) & (lin)  & (quad) & (lin)   & (quad)   \\
\hline
$\mathcal{P}_1$   & $1.4662$ & $0.9434$ & $2.8494$  & $0.9739$ & $0.9973$ & $0.8236$ & $0.5118$ & $29.6773$ & $18.0106$  \\
$\mathcal{P}_2$   & $1.5979$ & $1.3786$ & $3.8006$  & $0.9697$ & $1.0018$ & $0.8258$ & $-$ & $22.4069$ & $-$   \\
$\mathcal{P}_3$   & $1.4662$ & $0.9434$ & $2.8494$  & $0.5397$ & $0.3126$ & $0.9858$ & $0.9352$ & $64.1009$ & $104.9981$ \\
$\mathcal{P}_4$   & $1.5979$ & $1.3786$ & $3.8006$  & $0.5097$ & $0.2435$ & $0.9609$ & $0.9339$ & $49.6020$ & $100.9129$ \\
$\mathcal{P}_5$   & $1.5413$ & $4.8539$ & $9.0227$  & $1.0229$ & $0.9963$ & $-$ & $0.5143$ & $-$  & $5.7214$   \\
$\mathcal{P}_6$   & $1.6730$ & $4.9656$ & $9.9802$  & $1.0182$ & $0.9840$ & $-$ & $0.5513$ & $-$  & $5.6135$   \\
$\mathcal{P}_7$   & $1.5413$ & $4.8539$ & $9.0227$  & $0.3634$ & $0.1848$ & $1.0000$ & $0.9994$ & $30.4985$ & $59.9354$  \\
$\mathcal{P}_8$   & $1.6730$ & $4.9656$ & $9.9802$  & $0.2815$ & $0.0992$ & $1.0000$ & $0.9981$ & $35.5945$ & $100.8167$ \\
$\mathcal{P}_1^E$ & $2.9293$ & $1.6361$ & $7.7220$  & $0.8613$ & $1.0640$ & $0.9398$ & $-$ & $14.1304$ & $-$   \\
$\mathcal{P}_2^E$ & $3.0609$ & $1.8624$ & $8.7614$  & $0.8586$ & $1.0737$ & $0.9407$ & $-$ & 1$2.5056$ & $-$   \\
$\mathcal{P}_3^E$ & $2.9293$ & $1.6361$ & $7.7220$  & $0.3569$ & $0.1784$ & $0.8948$ & $0.8941$ & $32.4690$ & $64.9040$  \\
$\mathcal{P}_4^E$ & $3.0609$ & $1.8624$ & $8.7614$  & $0.3623$ & $0.1820$ & $0.8951$ & $0.8941$ & $28.2002$ & $56.0726$  \\
$\mathcal{P}_5^E$ & $3.0806$ & $5.5470$ & $20.1687$ & $0.9973$ & $0.9471$ & $0.5086$ & $0.6280$ & $2.5286$  & $3.2875$   \\
$\mathcal{P}_6^E$ & $3.2122$ & $5.6044$ & $21.2147$ & $0.9929$ & $0.9363$ & $0.5217$ & $0.6522$ & $2.4768$  & $3.2836$   \\
$\mathcal{P}_7^E$ & $3.0806$ & $5.5470$ & $20.1687$ & $0.1170$ & $0.1240$ & $0.9915$ & $0.9780$ & $42.0164$ & $39.1058$  \\
$\mathcal{P}_8^E$ & $3.2122$ & $5.6044$ & $21.2147$ & $0.1527$ & $0.1583$ & $0.9920$ & $0.9736$ & $30.6233$ & $28.9921$ \\
\hline
\end{tabular}
\caption{Resource and mitigation efficiency values of pipelines on \texttt{ibm\_lagos} with local folding \texttt{ZNE}. Some pipelines do not generate significantly higher proportions of \texttt{SUCCESS}, and those pipelines are marked with a dash ($-$) in the $PSR$ columns. Again, these pipelines reiterate the importance of statistical testing for pipelines. By assigning confidence intervals to the proportion of \texttt{SUCCESS}, we can parameterize the uncertainty of pipelines and understand which pipelines can consistently mitigate errors (and which ones cannot).}
\label{tab:lagos-local}
\end{table}
\endgroup

% perth local
\FloatBarrier
\begingroup
\renewcommand{\arraystretch}{1.5} %
\begin{table}
\centering
\begin{tabular}{cccccccccc}
\hline 
Pipeline          & T      & S      & R       & REM    & REM    & PSR    & PSR    & M       & M        \\
                  &        &        &         & (lin)  & (quad) & (lin)  & (quad) & (lin)   & (quad)   \\
\hline
$\mathcal{P}_1$   & $1.1863$ & $0.9439$ & $2.3061$  & $0.9785$ & $0.8899$ & $0.7251$ & $0.8897$ & $32.1352$ & $43.3540$  \\
$\mathcal{P}_2$   & $1.2998$ & $1.3996$ & $3.1189$  & $0.9657$ & $0.8624$ & $0.7308$ & $0.8741$ & $24.2635$ & $32.4991$  \\
$\mathcal{P}_3$   & $1.1863$ & $0.9439$ & $2.3061$  & $0.6899$ & $0.4267$ & $0.9623$ & $0.9948$ & $60.4875$ & $101.0925$ \\
$\mathcal{P}_4$   & $1.2998$ & $1.3996$ & $3.1189$  & $0.5950$ & $0.3007$ & $0.9055$ & $0.9811$ & $48.7956$ & $104.6124$ \\
$\mathcal{P}_5$   & $1.2613$ & $4.8542$ & $7.3839$  & $0.9150$ & $0.7643$ & $0.5868$ & $0.8701$ & $8.6850$  & $15.4168$  \\
$\mathcal{P}_6$   & $1.3747$ & $4.9673$ & $8.2034$  & $0.8683$ & $0.6633$ & $0.6648$ & $0.8893$ & $9.3333$  & $16.3435$  \\
$\mathcal{P}_7$   & $1.2613$ & $4.8542$ & $7.3839$  & $0.5947$ & $0.4376$ & $0.9918$ & $0.9878$ & $22.5863$ & $30.5717$  \\
$\mathcal{P}_8$   & $1.3747$ & $4.9673$ & $8.2034$  & $0.4860$ & $0.2973$ & $0.9955$ & $0.9777$ & $24.9685$ & $40.0875$  \\
$\mathcal{P}_1^E$ & $2.3695$ & $1.6365$ & $6.2473$  & $0.8004$ & $0.7445$ & $0.8633$ & $0.7506$ & $17.2650$ & $16.1384$  \\
$\mathcal{P}_2^E$ & $2.4829$ & $1.8740$ & $7.1360$  & $0.7818$ & $0.7673$ & $0.8715$ & $0.7237$ & $15.6210$ & $13.2176$  \\
$\mathcal{P}_3^E$ & $2.3695$ & $1.6365$ & $6.2473$  & $0.4381$ & $0.6286$ & $0.8072$ & $0.7194$ & $29.4923$ & $18.3186$  \\
$\mathcal{P}_4^E$ & $2.4829$ & $1.8740$ & $7.1360$  & $0.4860$ & $0.7369$ & $0.7951$ & $0.6646$ & $22.9262$ & $12.6388$  \\
$\mathcal{P}_5^E$ & $2.5204$ & $5.5473$ & $16.5019$ & $0.6751$ & $0.4020$ & $0.8575$ & $0.8498$ & $7.6971$  & $12.8107$  \\
$\mathcal{P}_6^E$ & $2.6339$ & $5.6053$ & $17.3975$ & $0.6288$ & $0.3678$ & $0.8535$ & $0.8553$ & $7.8020$  & $13.3672$  \\
$\mathcal{P}_7^E$ & $2.5204$ & $5.5473$ & $16.5019$ & $0.1578$ & $0.1625$ & $0.9258$ & $0.9363$ & $35.5540$ & $34.9154$  \\
$\mathcal{P}_8^E$ & $2.6339$ & $5.6053$ & $17.3975$ & $0.2470$ & $0.2449$ & $0.9332$ & $0.9322$ & $21.7159$ & $21.8802$\\
\hline
\end{tabular}
\caption{Resource and mitigation efficiency values of pipelines on \texttt{ibm\_perth} with local folding \texttt{ZNE}.}
\label{tab:perth-local}
\end{table}
\endgroup
\FloatBarrier

\bibliographystyle{unsrt} 
\bibliography{references}

\newpage
\appendix

\section{\texttt{ZNE} with Global Folding}\label{sec:global-folding-apndx}
\begin{figure}[t]
    \centering
    \includegraphics[width=6.4in]{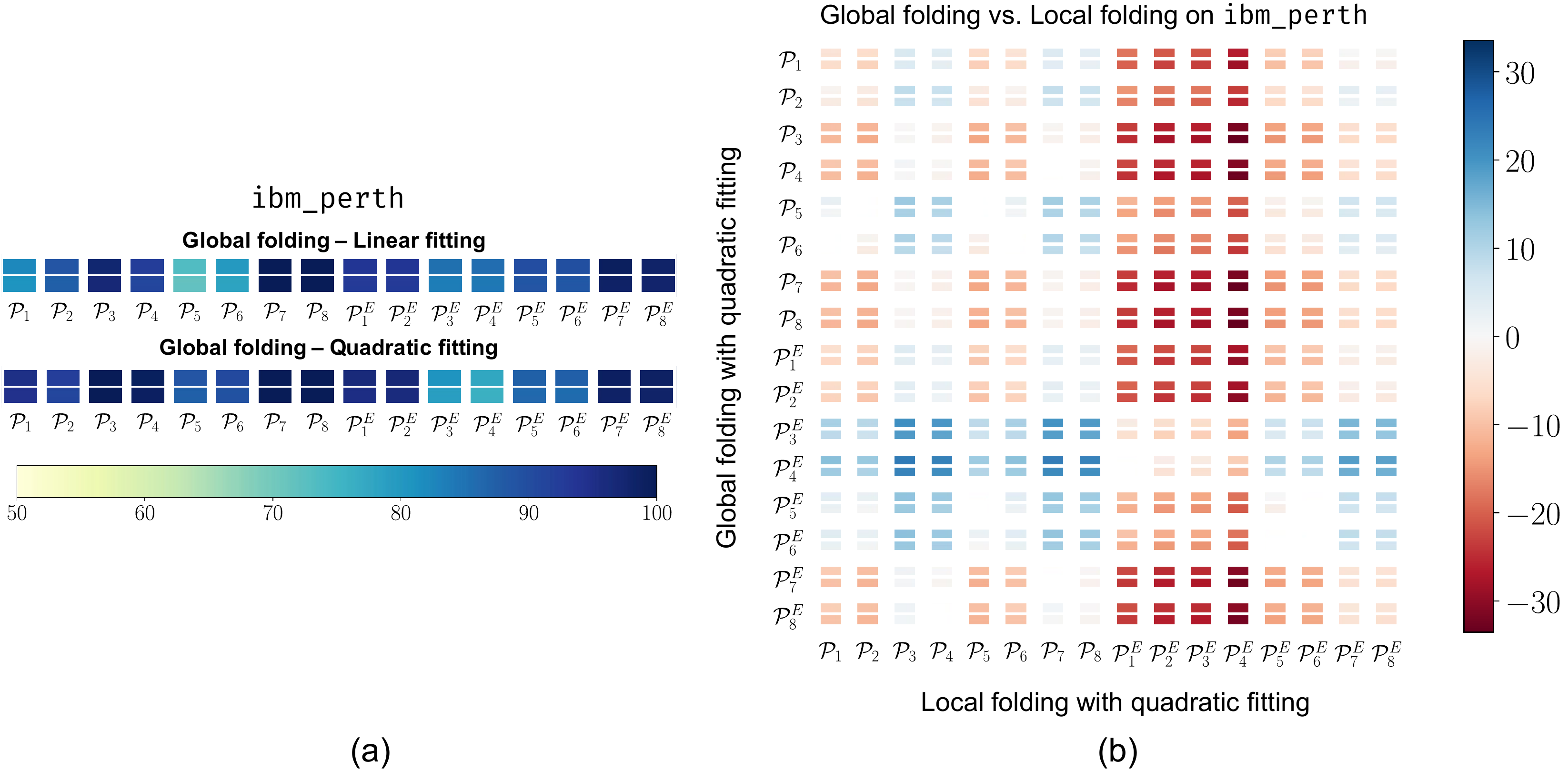}
    \caption{(a) Confidence intervals of proportion of \texttt{SUCCESS} for different EM pipelines with global folding \texttt{ZNE}.
    (b) Confidence intervals of comparison (difference) of proportion of \texttt{SUCCESS} between pipelines with global folding \texttt{ZNE} (quadratic fitting) (Y-axis) and local folding \texttt{ZNE} (quadratic fitting) (X-axis) on \texttt{ibm\_perth}. Pipelines $\mathcal{P}_7$ and $\mathcal{P}_8$ with global folding have higher proportions of \texttt{SUCCESS} than all the pipelines with local folding.
    }
    \label{fig:one-sample-conf-int-perth-global}
\end{figure}

We run the $16$ EM pipelines with the variant of \texttt{ZNE} with global folding on \texttt{ibm\_perth}. We evaluate the individual performance of global folding \texttt{ZNE} and compare it with \texttt{ZNE} with local (CNOT) folding. Figure~\ref{fig:one-sample-conf-int-perth-global}a shows the $95\%$ confidence intervals of proportion of \texttt{SUCCESS} EM pipelines with global folding \texttt{ZNE}. All pipelines, for both linear and quadratic fitting, generate a significantly higher proportion of \texttt{SUCCESS} than the proportion of \texttt{FAIL}, as indicated by all the blue confidence intervals. Similarly to the local folding case, pipelines $\mathcal{P}_3$, $\mathcal{P}_4$, $\mathcal{P}_7$, $\mathcal{P}_8$, $\mathcal{P}_7^E$, and $\mathcal{P}_8^E$ generate the darkest blue intervals for global folding as well. 

Next, we compare the performance of global folding (quadratic fitting) with local folding (quadratic fitting) to infer if one is more successful than the other and, hence, a possible better choice. The comparison is plotted in Figure~\ref{fig:one-sample-conf-int-perth-global}b. The heat map shows the confidence intervals of the proportion difference of \texttt{SUCCESS} between linear and global folding. As most of the cells in the heat map are dimmed, we can infer that the difference between linear and global folding variants of \texttt{ZNE} is nuanced. For instance, global folding variants of $\mathcal{P}_7$ and $\mathcal{P}_8$ outperform every pipeline with local folding (as indicated by all-red rows). However, the proportion of \texttt{SUCCESS} of $\mathcal{P}_7$ and $\mathcal{P}_8$ (global) is marginally higher ($\approx 1\%-2\%$) than $\mathcal{P}_7$ and $\mathcal{P}_8$ (local). 

\begin{figure}[t]
    \centering
    \includegraphics[width=6in]{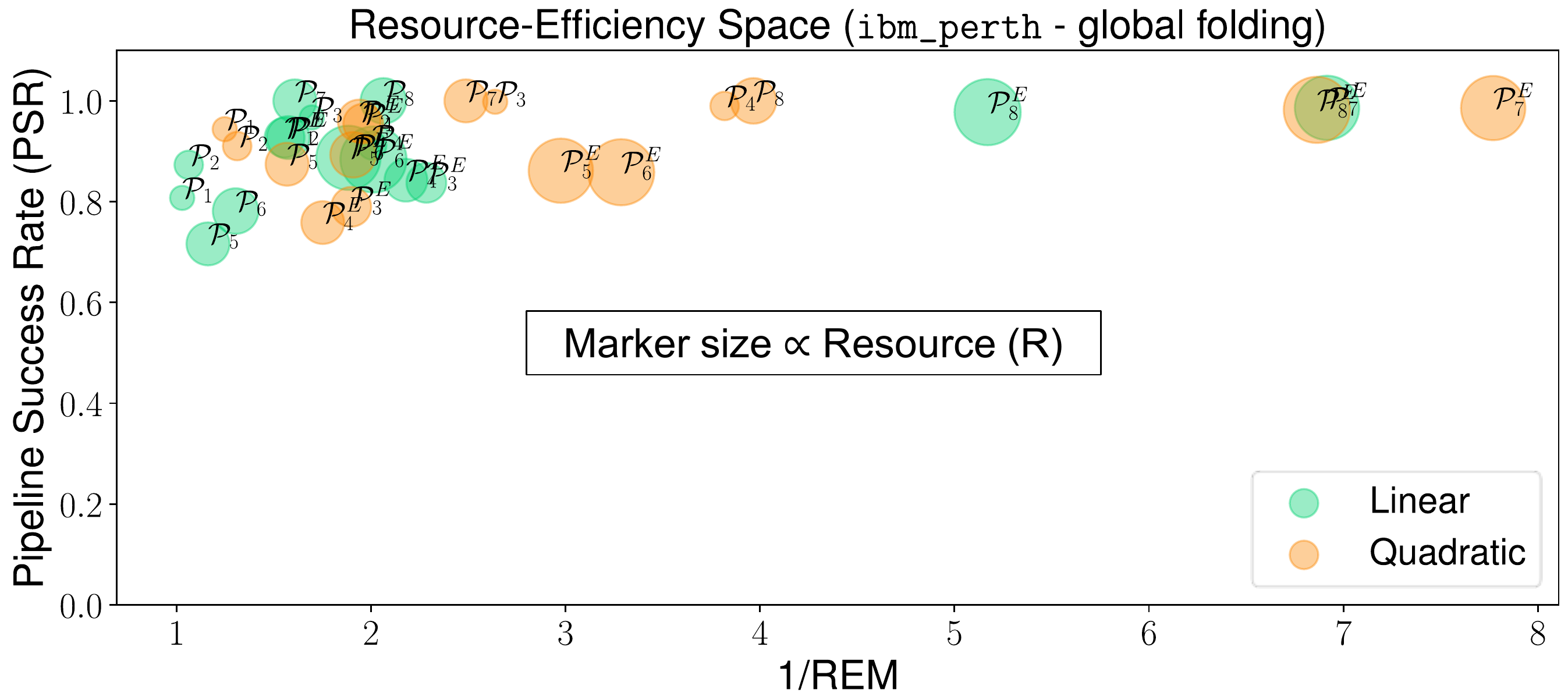}
    \caption{Resource-Efficiency space for experiments on \texttt{ibm\_perth} with global folding \texttt{ZNE}. $\mathcal{P}_7^E$ (quadratic fitting) stands out among others in terms of mitigation efficiency, albeit with a high resource consumption. $\mathcal{P}_4$ (quadratic fitting) provides a middle-ground between resource and efficiency. Many pipelines consume more resource than $\mathcal{P}_4$ (quadratic) while providing inferior mitigation. Thus, by selecting an EM pipeline intelligently users can save resources.}
    \label{fig:resource-vs-accuracy-perth-global}
\end{figure}

Finally, we plot the \emph{resource-efficiency space} for global folding experiments in Figure~\ref{fig:resource-vs-accuracy-perth-global}. The complete data is tabulated in Table~\ref{tab:resource-performance-perth-global}. $\mathcal{P}_7^E$ (quadratic) is a standout among other pipelines. It has a high $PSR$ ($98\%$) and the lowest $REM$ ($0.1287$). It performs better than its local folding counterpart in terms of both $PSR$ and $REM$. While it is the most efficient pipeline in mitigating errors, $\mathcal{P}_7^E$ is the second most resource-intensive pipeline. 
$\mathcal{P}_4$ and $\mathcal{P}_8$ (quadratic) provide a decent middle-ground between resource and efficiency. The pipeline has high $PSR$ ($99\%$), moderately low $REM$ ($0.26$), and second lowest resource usage.


\begingroup
\renewcommand{\arraystretch}{1.5} %

\begin{table}[H]
\centering
\begin{tabular}{cccccccccc}
\hline
Pipeline          & \multicolumn{1}{c}{T} & \multicolumn{1}{c}{S} & \multicolumn{1}{c}{R} & \multicolumn{1}{c}{REM}   & \multicolumn{1}{c}{REM}    & \multicolumn{1}{c}{PSR}   & \multicolumn{1}{c}{PSR}    & \multicolumn{1}{c}{M}     & \multicolumn{1}{c}{M}      \\
                  & \multicolumn{1}{c}{}  & \multicolumn{1}{c}{}  & \multicolumn{1}{c}{}  & \multicolumn{1}{c}{(lin)} & \multicolumn{1}{c}{(quad)} & \multicolumn{1}{c}{(lin)} & \multicolumn{1}{c}{(quad)} & \multicolumn{1}{c}{(lin)} & \multicolumn{1}{c}{(quad)} \\
\hline
$\mathcal{P}_1$   & $1.0519$              & $0.9658$              & $2.0679$              & $0.9716$                  & $0.8013$                   & $0.8075$                  & $0.9439$                   & $40.1902$                 & $56.9615$                  \\
$\mathcal{P}_2$   & $1.1654$              & $1.4608$              & $2.8678$              & $0.9407$                  & $0.7622$                   & $0.8730$                  & $0.9104$                   & $32.3611$                 & $41.6480$                  \\
$\mathcal{P}_3$   & $1.0519$              & $0.9658$              & $2.0679$              & $0.5898$                  & $0.3790$                   & $0.9671$                  & $0.9983$                   & $79.2916$                 & $127.3710$                 \\
$\mathcal{P}_4$   & $1.1654$              & $1.4608$              & $2.8678$              & $0.4985$                  & $0.2619$                   & $0.9116$                  & $0.9899$                   & $63.7657$                 & $131.8003$                 \\
$\mathcal{P}_5$   & $1.1052$              & $4.8783$              & $6.4966$              & $0.8606$                  & $0.6374$                   & $0.7162$                  & $0.8743$                   & $12.8108$                 & $21.1147$                  \\
$\mathcal{P}_6$   & $1.2186$              & $4.9919$              & $7.3018$              & $0.7668$                  & $0.5241$                   & $0.7813$                  & $0.8930$                   & $13.9544$                 & $23.3348$                  \\
$\mathcal{P}_7$   & $1.1052$              & $4.8783$              & $6.4966$              & $0.6216$                  & $0.4019$                   & $1.0000$                  & $1.0000$                   & $24.7630$                 & $38.2997$                  \\
$\mathcal{P}_8$   & $1.2186$              & $4.9919$              & $7.3018$              & $0.4845$                  & $0.2521$                   & $1.0000$                  & $1.0000$                   & $28.2668$                 & $54.3247$                  \\
$\mathcal{P}_1^E$ & $2.0943$              & $1.6591$              & $5.5689$              & $0.6427$                  & $0.5101$                   & $0.9272$                  & $0.9549$                   & $25.9049$                 & $33.6152$                  \\
$\mathcal{P}_2^E$ & $2.2077$              & $1.9188$              & $6.4439$              & $0.6353$                  & $0.5156$                   & $0.9272$                  & $0.9600$                   & $22.6481$                 & $28.8952$                  \\
$\mathcal{P}_3^E$ & $2.0943$              & $1.6591$              & $5.5689$              & $0.4377$                  & $0.5268$                   & $0.8375$                  & $0.7897$                   & $34.3588$                 & $26.9191$                  \\
$\mathcal{P}_4^E$ & $2.2077$              & $1.9188$              & $6.4439$              & $0.4588$                  & $0.5708$                   & $0.8426$                  & $0.7585$                   & $28.5001$                 & $20.6219$                  \\
$\mathcal{P}_5^E$ & $2.2078$              & $5.5713$              & $14.5080$             & $0.5308$                  & $0.3360$                   & $0.8870$                  & $0.8615$                   & $11.5188$                 & $17.6723$                  \\
$\mathcal{P}_6^E$ & $2.3212$              & $5.6297$              & $15.3889$             & $0.4968$                  & $0.3043$                   & $0.8838$                  & $0.8584$                   & $11.5597$                 & $18.3308$                  \\
$\mathcal{P}_7^E$ & $2.2078$              & $5.5713$              & $14.5080$             & $0.1446$                  & $0.1287$                   & $0.9862$                  & $0.9858$                   & $47.0095$                 & $52.7939$                  \\
$\mathcal{P}_8^E$ & $2.3212$              & $5.6297$              & $15.3889$             & $0.1934$                  & $0.1457$                   & $0.9777$                  & $0.9822$                   & $32.8499$                 & $43.8053$    \\
\hline
\end{tabular}
\caption{Resource and mitigation efficiency values of pipelines on \texttt{ibm\_perth} with global folding \texttt{ZNE}.}
\label{tab:resource-performance-perth-global}
\end{table}
\endgroup

\end{document}